\documentclass[12pt]{article}

\usepackage{latexsym,amsmath,amssymb,amsthm,amsfonts,graphicx,natbib,breqn,epsfig,bm,authblk,url,thmtools}

\usepackage{float} 
\usepackage{booktabs} 
\usepackage{graphicx} 
\usepackage[margin=1cm]{caption} 
\usepackage{mathtools}
\usepackage{enumerate}
\usepackage{bbm}
\usepackage{tikz}
\usepackage[colorinlistoftodos]{todonotes}
\usetikzlibrary{shapes.multipart}
\usetikzlibrary{arrows, positioning}
\usepackage{algorithm}

\floatstyle{plain}
\newfloat{Algorithm}{thp}{lop}
\floatname{Algorithm}{Algorithm}

\declaretheoremstyle[
notefont=\bfseries, notebraces={}{},
bodyfont=\normalfont,
postheadspace=0.5em,
numbered=yes,
]{mystyle}

\theoremstyle{definition}

\begin{document}

\renewcommand{\thefootnote}{\fnsymbol{footnote}}

\title{\vspace{-10mm} Weakly informative priors and prior-data conflict checking for likelihood-free inference}

\author[1]{Atlanta Chakraborty}
\author[1,2]{David J. Nott\thanks{Corresponding author:  standj@nus.edu.sg}}
\author[3]{Michael Evans}
\affil[1]{Institute of Operations Research and Analytics,
National University of Singapore, Singapore 117602}
\affil[2]{Department of Statistics and Applied Probability,
 National University of Singapore, Singapore 117546}
\affil[3]{Department of Statistics,
University of Toronto, Toronto, ON M5S 3G3, Canada}

\date{}

\maketitle


\vspace*{-15mm}\noindent

\begin{abstract}
Bayesian likelihood-free inference, which is used to perform Bayesian inference when the likelihood is intractable, 
enjoys an increasing number of important scientific applications.  
However, many aspects of a Bayesian analysis become more challenging in the likelihood-free setting.  
One example of this is prior-data conflict checking, where the goal is to assess whether the
information in the data and the prior are inconsistent.  Conflicts of this kind
are important to detect, since they may reveal problems in an investigator's understanding of 
what are relevant values of the parameters, 
and can result in sensitivity of Bayesian inferences to the prior.  
Here we consider methods for prior-data conflict checking which are
applicable regardless of whether the likelihood is tractable or not.  
In constructing our checks, we consider checking statistics based on prior-to-posterior Kullback-Leibler divergences.
The checks are implemented using mixture approximations to the posterior distribution and closed-form
approximations to Kullback-Leibler divergences for mixtures, which make
Monte Carlo approximation of reference distributions for calibration computationally feasible.  
When prior-data conflicts occur, it is useful to
consider weakly informative prior specifications in alternative analyses as part of a 
sensitivity analysis.  As a main application of our methodology, 
we develop a technique for searching for weakly informative priors in likelihood-free inference, where
the notion of a weakly informative prior is formalized using prior-data conflict checks.  
The methods are demonstrated in three examples.  

\vspace{2mm}

\end{abstract}

\section{Introduction}

It is often natural to translate scientific knowledge into an appropriate statistical model through
specification of a generative process for the data, 
and this leads to models defined in terms of a simulation algorithm rather than through an explicit mathematical formulation.  
For these kinds of models, computation of the likelihood may be intractable, and then likelihood-free inference methods,
which simulate from the model as a surrogate for likelihood evaluations, can be used.  
Currently the two most popular Bayesian likelihood-free inference approaches are approximate Bayesian
computation (ABC) \citep{pritchard+spf99,beaumont+zb02,sisson+fb18} and synthetic likelihood \citep{wood10,price+dln16}, and the further development of these and other likelihood-free inference algorithms is an active topic of current
research.   The purpose of the current paper is to develop some tools for checking for prior-data conflict which
are applicable when the likelihood is intractable.  This means developing checks which can be computed using
only simulation from the model, without requiring evaluation of the likelihood.
As a main application of our methodology, a technique for searching for a weakly informative prior 
with respect to an elicited prior is also developed, where the notion of a weakly informative prior 
is formalized using prior-data conflict checks.

For complex models, a challenging aspect of any Bayesian analysis is specification of the prior distribution,
since an inadequate elicitation process may result in a prior distribution that is informative in ways that are unintended.   
If an informative prior has been used,  
one approach to guarding against undesirable prior sensitivity
is to check for the existence of prior-data conflicts, which occur
when the prior puts all its mass out in the tails of the likelihood.  
Prior-data conflicts are important to detect, since they indicate a lack of understanding in setting
up the model.  Furthermore, prior sensitivity of inferences will increase with the severity
of the conflict \citep{allabadi+e17}.  A difficulty with many prior-data conflict checking methods, however, is that 
the required computations are demanding, even when the likelihood is tractable.  

It is especially important
in the context of Bayesian likelihood-free inference to develop prior-data conflict checking methods, 
since alternative techniques for investigating prior sensitivity or exploring conflicts are usually unavailable.  For example, 
objective Bayes methods \citep{berger+bs09} which specify a prior as a reference for comparison
usually cannot be implemented, since determining these involves computations using the likelihood.
Here we develop an approach to prior-data conflict checking which is
applicable whether the likelihood is tractable or not. 
We consider the conflict checks recently suggested in \cite{nott+wee16}, which
use prior-to-posterior divergences as checking statistics.  To make computations tractable, we use
mixture approximations to the posterior distribution, which makes repeated computations
of posterior distributions for different datasets feasible.  These together with closed form
approximations of the Kullback-Leibler divergence for mixtures can be used to calculate
tail probabilities for calibration of the checks in a computationally tractable way.  

When prior-data conflicts occur, it can be helpful to consider an alternative analysis using a weakly informative
prior which retains some of the original prior information but resolves the conflict, in order to see how this affects
conclusions of interest.  
\cite{evans+j11}, inspired by \cite{gelman06}, developed a formalization of the notion of a weakly informative
prior relative to a base prior which uses a prior-data conflict check in the definition. 
As a main application of our methodology, we develop  
convenient methods for searching for weakly
informative priors in the sense of \cite{evans+j11}.    
While these weakly informative priors are a useful
tool for exploring prior sensitivity, the goals of prior-data conflict checking and development of associated
weakly informative priors do not relate solely to Bayesian sensitivity analysis, 
for which there is a large existing literature (\cite{mcculloch89}, \cite{lavine91}, \cite{clarke+g98}, 
\cite{zhu+it11}, \cite{roos+mhr15}, 
among many others). 
See \cite{allabadi+e17} for further discussion of the relationship between
prior sensitivity and prior-data conflict.

In the next section we give an introduction to some of the existing literature on Bayesian model checking, and consider
in some detail the proposal of \cite{nott+wee16} for prior-data conflict checks based on prior-to-posterior
divergences.  We also develop an implementation of this procedure for the likelihood-free case, based on 
mixture posterior approximations and closed-form approximations to Kullback-Leibler divergences
for mixtures. Similar approximate checks
were considered in \cite{nott+wee16} for the case of a tractable likelihood where 
mixture variational approximations were used for posterior computations.  Because their variational
approximation methods
require evaluations of the likelihood, they do not apply in the likelihood-free setting.  
Hence, mixture approximations need to be obtained in a different way
in the case of an intractable likelihood, and that is achieved here by fitting mixture models
to approximate the joint density of summary statistics and model parameters.  Once the approximation
to the joint density is obtained, approximations to the posterior density for the parameters
given summary statistics can be induced for different values of the summary statistics 
at negligible additional computational cost.  
This is crucial to the computational tractability of our approach to searching for weakly informative priors, which
is described in Section 3.
Section 4 considers a number of examples and Section 5 gives some concluding
discussion.

\section{Prior-data conflict checking}

\subsection{Basic ideas of prior-data conflict checking}

Let $\theta$ be a parameter, $y$ be data, $p(\theta)$ be a prior density for $\theta$, $p(y|\theta)$ be the sampling
density for $y$ given $\theta$ and $p(\theta|y)$ be the posterior density.  In a Bayesian analysis, prior-data conflict
occurs when the prior density puts all its mass out in the tails of the likelihood, so that the information in the data about
$\theta$ and the information in the prior are in conflict.  Various methods have been developed for
checking for prior-data conflict (\cite{ohagan03,marshall+s07,evans+m06,gasemyr+n09,evans+j10,presanis+osd13,nott+wee16}, among many
others).  However, many of these methods are difficult to apply in the case of a model with an intractable likelihood.
A prior-data conflict checking method is applicable with intractable likelihood if the check can be conducted using
only simulation of data from the model, without evaluation of the likelihood.  One method
that can be applied in a likelihood-free setting is described in
\cite{nott+dme18} who considered a certain implementation of the approach of \cite{evans+m06}.  
However, the method of \cite{nott+dme18} relies on kernel density estimation of a vector summary statistic, 
which is difficult when the dimension of the summary statistic is moderately large.   The method of 
\cite{evans+m06} also lacks a desirable parametrization invariance property in the case of a continuous parameter
where the check can depend on the choice of sufficient statistic.  Further discussion of the statistical
properties of the checks of \cite{nott+wee16} and \cite{evans+m06}, which are the basis for the likelihood-free versions
of those checks in the present work and in \cite{nott+dme18} respectively, is given in \cite{nott+wee16}.

A prior-data conflict check is a special kind of Bayesian predictive check of the kind used for Bayesian model criticism.  See, 
for example, \cite{gelman+ms96}, \cite{bayarri+c07} and \cite{evans15} 
for general overviews of Bayesian model checking.  
A Bayesian predictive check involves the choice of a statistic and reference distribution.  Write $T=T(y)$ for
a scalar statistic, and suppose that we wish to criticize the model by determining whether
the observed value $t_{\text{obs}}$ of $T$ is surprising under some reference distribution $m(t)$.  As a measure
of surprise, a Bayesian predictive
$p$-value can be computed as 
\begin{align}
  p & = P(T\geq t_{\text{obs}}),  \label{bpcheck}
\end{align}
where $T\sim m(t)$ and it has been assumed above that $T$ is defined in such a way that a large value 
indicates a possible model failure.  Note that the purpose of (\ref{bpcheck}) is to locate where
$t_{\text{obs}}$ lies with respect to the distribution of $T$.  
\cite{evans+m06} consider the question of what are logical requirements on the
statistic $T$ and the reference distribution $m(t)$ when the goal is to check for prior-data conflict.  They answer
this question by generalizing a decomposition of the joint model for $(y,\theta)$ due to \cite{box80}, and 
consider the terms in the decomposition as playing different roles in the analysis.  For prior-data conflict checks, 
$T$ plays the role of summarizing the likelihood, and $T$ should not depend on aspects of $y$ 
that are irrelevant to the likelihood;  this means
that $T$ should be a function of a minimal sufficient statistic.  Furthermore, any check based on a $T$ which is a function
of a minimal sufficient statistic should be invariant to the minimal sufficient statistic chosen.  For detecting an inconsistency between the likelihood and prior, we want to see whether the observed
likelihood (summarized by the observed value $t_{\text{obs}}$ of $T$) is unusual compared to what is expected
under the prior.  This means that the reference distribution $m(t)$ should be the prior predictive distribution of
$T$, which we write as $p(t)=\int p(t|\theta)p(\theta)\,d\theta$, where $p(t|\theta)$ denotes the sampling distribution
of $T$ given $\theta$.  

The prior-data conflict checks considered in \cite{evans+m06} are not invariant to the choice of minimal sufficient
statistic, and a modified version which is invariant but difficult to apply is discussed in \cite{evans+j10}.  
\cite{evans+m06}
also consider conditioning on ancillary statistics, and extensions to separately checking components of 
hierarchical priors, but we do not consider this further here.  
One way to obtain a statistic that is a function of any sufficient statistic and invariant to its choice is to consider
some function of the posterior distribution itself. \cite{nott+wee16} consider an approach of this kind, where
the statistic $T$ is a prior-to-posterior R\'{e}nyi divergence, and 
it is a further development of this approach that is the focus of the current work.

\subsection{Conflict checks using prior-to-posterior divergence}

The prior-data conflict checks of \cite{nott+wee16} use a prior-to-posterior R\'{e}nyi divergence as the 
checking statistic.  Here we consider the special case of the Kullback-Leibler divergence,
resulting in the checking statistic
\begin{align}
  G & = \text{KL}(p(\theta|y)|| p(\theta)) \nonumber \\
   & \overset{\operatorname{\text{def}}}{=} \int \log \frac{p(\theta|y)}{p(\theta)} p(\theta|y)\,d\theta.  \label{prior-to-posterior}
\end{align}
To calibrate the observed value of this statistic we use a tail probability (Bayesian predictive $p$-value)
\begin{align}
  p_{\text{KL}} & = P(G\geq G_{\text{obs}}),  \label{prior-to-posterior-tail}
\end{align}
where $G\sim p(g)$ with $p(g)$ the prior-predictive density of $G$, and $G_{\text{obs}}$ denotes the observed value.  
It is possible in principle to replace the Kullback-Leibler divergence with other divergences in the check (\ref{prior-to-posterior}), 
but using the Kullback-Leibler divergence is convenient computationally here, allowing us to make use of 
closed-form approximations for Kullback-Leibler divergences between Gaussian mixture distributions.  This is described
later and allows
approximate versions of the check (\ref{prior-to-posterior}) to be implemented rapidly, which is particularly important
in our application to searching for weakly informative priors.  

If we are to use the above check in likelihood-free inference problems, we need to implement it using only
simulation from the model, without requiring evaluation of the likelihood.  Before we describe how this can
be done, however, it is useful to give some context about why likelihood-free inference is used.
The earliest applications of likelihood-free inference arose in population genetics in the
form of ABC algorithms \citep{pritchard+spf99}, but these and similar methods are now used in a wide range of problems
where the likelihood is intractable
due to complex observation models or difficulty in integrating out complex latent processes.  
There are other more specific motivations in particular applications.  For example, 
in developing the
synthetic likelihood method, \cite{wood10}  
considered time series models for ecological data with chaotic dynamics and low enviornmental noise.
In these models the likelihood may be difficult to evaluate using methods relying on state estimation for
state space models -- see \cite{fasiolo+pw16} for further elaboration and Section 4.3 for an example of this kind considered
in \cite{fasiolo+whb18}.  
Another motivation for using likelihood-free methods is to robustify Bayesian analyses with tractable likelihood
by basing information only on (possibly complex) summary statistics.  The summary statistic likelihood is often intractable,
but considering an insufficient statistic which discards information can be useful in the case of misspecified models -- see
\cite{lewis+ml21} for a recent discussion of the statistical motivation here, although the authors focus on applications
to linear models and do not use likelihood-free methods for computation.  \cite{sisson+fb18} is a recent comprehensive overview
of likelihood-free inference methods discussing a wide range of methods and applications.  

To implement a check based on the statistic (\ref{prior-to-posterior}) in the likelihood-free setting, 
we make several approximations.  The first is to consider replacing the posterior
distribution $p(\theta|y)$ with the posterior distribution given a summary statistic, say $z=z(y)$ in (\ref{prior-to-posterior}).   
Most likelihood-free inference methods, such as ABC and synthetic likelihood, make use of reduced dimension 
summary statistics for the data
since they use empirical methods based on simulated data to estimate the distribution
of the summary statistics for likelihood estimation.  For example, the ABC approach can be regarded as estimating
the likelihood based on a kernel density estimate of the summary statistic density, and there is a curse of dimensionality
associated with the use of kernel methods, so that a low-dimensional summary statistic is desirable.  
Ideally the summary statistic is sufficient, so that no
information about $\theta$ is lost, but non-trivial sufficient summary statistics will not usually be available.  
See \cite{blum+nps13} and \cite{prangle18} for further discussion of the issue of summary statistic 
choice in likelihood-free inference.  

The dimension reduction achieved by using summary statistics is useful for implementing our next approximation, 
which is to use a mixture model to estimate the posterior distribution of the parameters given summary statistic
values.  Mixture approximations have been used in the ABC context before.  For example, \cite{bonassi+yw11} consider
mixture modelling of parameter and summary statistics jointly and the induced conditional distribution for the parameters
as a form of nonlinear regression adjustment.  \cite{bonassi+w15} consider
similar mixture approximations within sequential Monte Carlo ABC schemes, and \cite{fan+ns13} consider
an approach to estimating the likelihood using mixtures of experts and copulas.  
\cite{forbes+nna21} use mixture of experts approximations to the posterior distribution directly, and 
use their mixture estimates to define discrepancy measures in distribution space for ABC analyses.  \cite{he+hy21}
have recently considered variational approximation of the posterior density using a mixture family in likelihood-free
inference problems.
The method considered below is the method considered in \cite{bonassi+yw11}.  
The great advantage of this approach here is that it can allow us to produce repeated posterior approximations
for different data at low computational cost, which is important for approximating the reference distribution
of the conflict check in computing (\ref{prior-to-posterior}).  This is also important in the application of our checks
to searching for weakly informative priors in the next section.  

The mixture approximations we consider are obtained in the following way.  
Write $x=(\theta,z)$,  and suppose we sample parameter value and summary statistic pairs
$x_i=(\theta_i,z_i)$, $i=1,\dots, n$, 
from $p(x)=p(\theta,z)=p(\theta)p(z|\theta)$.   
The posterior density of $\theta$ given $z_{\text{obs}}$ is the conditional density of $\theta$ given $z=z_{\text{obs}}$
derived from the joint density $p(x)=p(\theta,z)$.  We fit a Gaussian mixture model to $x_i$, $i=1,\dots, n$, to obtain a Gaussian
mixture approximation to $p(\theta,z)$, which we denote by $\widetilde{p}(x)$, 
\begin{align}
 \widetilde{p}(x) & = \sum_{j=1}^{J} w_j \phi_j(x), \label{mixture}
\end{align}
where $J$ is the number of mixture components, $w_j$ are non-negative mixing weights summing
to one, and $\phi_j(x)=\phi(x;\mu_j,\Sigma_j)$ denotes a multivariate Gaussian density with mean vector $\mu_j$ and
covariance matrix $\Sigma_j$.  For a Gaussian mixture model, conditional distributions
are also Gaussian mixture models having easily computed closed form expressions.  
So once the joint density $p(x)$ has been approximated by $\widetilde{p}(x)$, we can obtain
the conditional density for $\theta$ given $z$, which
we denote by $\widetilde{p}(\theta|z)$.   To give an expression for this we need some
further notation.  Suppose we partition $\mu_j$ and $\Sigma_j$ in
the same way as $x=(\theta,z)$ as $\mu_j=(\mu_{j,\theta},\mu_{j,z})$ and 
\begin{align*}
  \Sigma_j=\left[\begin{array}{cc} \Sigma_{j,\theta} & \Sigma_{j,\theta z} \\  \Sigma_{j,z\theta} & \Sigma_{j,z} \end{array}\right].
\end{align*}
Then 
\begin{align}
  \widetilde{p}(\theta|z) = \sum_{j=1}^J w_{j|z}\phi_{j|z}(\theta),  \label{mixture-conditional}
\end{align}
where
$\phi_{j|z}(\theta)=\phi(\theta;\mu_{j|z},\Sigma_{j|z})$, with 
\begin{align*}
  \mu_{j|z} & = \mu_{j,\theta}+\Sigma_{j,\theta z}\Sigma_{j,z}^{-1}(z-\mu_{j,z}),
\end{align*}
\begin{align*}
  \Sigma_{j|z} & = \Sigma_{j,\theta}-\Sigma_{j,\theta z}\Sigma_{j,z}^{-1}\Sigma_{j,z \theta},
\end{align*}
and
\begin{align*}
  w_{j|z} & = \frac{w_j \phi_j(z)}{\sum_{l=1}^J w_l \phi_l(z)},
\end{align*}
where $\phi_j(z)=\phi(z;\mu_{j,z},\Sigma_{j,z})$.

The conditional density (\ref{mixture-conditional}) is an approximation to the posterior density of $\theta$ given $z$, and 
is easily computable for any summary statistic value $z$.  This is important since Monte Carlo approximation of the tail
probability (\ref{prior-to-posterior-tail}) involves approximating the posterior density repeatedly for different data.  
To approximate
(\ref{prior-to-posterior-tail}) using Monte Carlo, we generate summary statistic values $z^{(1)},\dots, z^{(R)}$ from the prior
predictive for $z$, then compute the approximate posterior densities
$\widetilde{p}(\theta|z_{\text{obs}})$ and $\widetilde{p}(\theta|z^{(r)})$, $r=1,\dots, R$, where 
$z_{\text{obs}}$ is the observed value for $z$.  
If we were able to compute the prior-to-posterior Kullback-Leibler divergences for our approximations, we would then 
compute the proportion of the simulated summary statistics for which the divergence was larger than that for the observed
summary statistic as in (\ref{prior-to-posterior-tail}).

To overcome the difficulty of computing the prior-to-posterior Kullback-Leibler divergence, we exploit
the fact that our posterior approximations are Gaussian mixtures, and assume that the prior can
be approximated as a Gaussian mixture also.  We write $\widetilde{p}(\theta)$ for the mixture
approximation to the prior.  If the prior is Gaussian or a Gaussian mixture, then $\widetilde{p}(\theta)=p(\theta)$, but
if it is not we might simulate
samples from the prior and then fit a mixture model as described to obtain $\widetilde{p}(\theta)$.  
A closed-form approximation for the Kullback-Leibler
divergence between two mixture models, due to \citet[Section 7]{hershey+o07}, is then used 
as in \cite{nott+wee16}.  
For this consider two mixture densities
$f(\theta)$ and $g(\theta)$,
$$f(\theta)=\sum_{j=1}^{J_f} w_{f,j}\phi_{f,j}(\theta),\;\;\; g(\theta)=\sum_{j=1}^{J_g} w_{g,j}\phi_{g,j}(\theta),$$
where $J_f$ and $J_g$ are the number of mixture components for $f$ and $g$ respectively, $w_{f,j}$, $j=1,\dots, J_f$ 
and $w_{g,j}$, $j=1,\dots, J_g$ are non-negative mixing weights for the respective densities summing to one, and 
$\phi_{f,j}(\theta)=\phi(\theta;\mu_{f,j},\Sigma_{f,j})$ and $\phi_{g,j}(\theta)=\phi(\theta;\mu_{g,j},\Sigma_{g,j})$
are respective multivariate normal component densities.  Then approximate the Kullback-Leibler divergence
$\text{KL}(g(\theta)||f(\theta))$ by 
\begin{align}
 \widetilde{\text{KL}}(g(\theta)||f(\theta)) & = \sum_{j=1}^{J_g} w_{g,j}\log \frac{\sum_{k=1}^{J_g} w_{g,k} \exp(-\text{KL}(\phi_{g,j}||\phi_{g,k})}
  {\sum_{l=1}^{J_f} w_{f,l}\exp(-\text{KL}(\phi_{g,j}||\phi_{f,l}))}, \label{KL-approx}
\end{align}
where the Kullback-Leibler divergences on the right-hand side 
in the above expression are between multivariate normal components densities, for which
there is an exact closed-form expression.  

Combining our normal mixture approximations to the prior and posterior and the approximation (\ref{KL-approx}), 
an approximate version of the prior-to-posterior Kullback-Leibler divergence statistic (\ref{prior-to-posterior}) for $z$ is
then given by
\begin{align}
\widetilde{G} & =\widetilde{G}(z)=\widetilde{\text{KL}}(\widetilde{p}(\theta|z)||\widetilde{p}(\theta)). \label{prior-to-posterior-approx}
\end{align}
Then our prior-data conflict checks for likelihood-free inference 
approximates (\ref{prior-to-posterior-tail}) by
\begin{align}
  \widetilde{p}_{\text{KL}} & =\frac{1}{R} \sum_{r=1}^R I(\widetilde{G}^{(r)}\geq \widetilde{G}_{\text{obs}}), \label{prior-to-posterior-tail-approx}
\end{align}
where $\widetilde{G}^{(r)}=\widetilde{G}(z^{(r)})$, $r=1,\dots, R$ are values of $\widetilde{G}$ for independent simulations
$z^{(r)}$, $r=1,\dots, R$, from the prior predictive distribution of $z$, 
$\widetilde{G}_{\text{obs}}=\widetilde{G}(z_{\text{obs}})$ is 
the value of $\widetilde{G}$ for the observed summary statistic value $z_{\text{obs}}$, and $I(\cdot)$ denotes the indicator
function.   The computations required for our conflict check are summarized in Algorithm \ref{approx-check}.

 \begin{algorithm}[!h] 
\caption{Computation of prior-data conflict check}
\label{approx-check}
   \vspace{0.1in}
  \noindent {\it Inputs:} 
  
  \begin{itemize}
  \item Prior distribution $p(\theta)$, model $p(z|\theta)$ for summary statistics $z$, observed summary statistic value
  $z_{\text{obs}}$.   
  \item Training sample size $n$ for fitting mixture approximation, number of replicates $R$ for Monte Carlo approximation of
  $p$-value.
  \end{itemize}
  
  \noindent{\it Output:}
  \begin{itemize}
  \item Tail probability $\widetilde{p}_{\text{KL}}$ given in (\ref{prior-to-posterior-tail-approx}).
  \end{itemize}
  
  \noindent{\it Initialization:}
  \begin{itemize}
  \item Simulate $x_i=(\theta_i,z_i) \sim p(x)$, $i=1,\dots, n$, and obtain a Gaussian mixture model approximation
  $\widetilde{p}(x)$ of $p(x)$. 
  \item If the prior $p(\theta)$ is not Gaussian or a Gaussian mixture, obtain a Gaussian mixture approximation
  $\widetilde{p}(\theta)$ of $p(\theta)$ by fitting to the samples $\theta_i$, $i=1,\dots, n$.  
  \end{itemize}
  
  \noindent{\it Computation of tail probability $\widetilde{p}_{\text{KL}}$:}
  \begin{enumerate}
  \item For $r=1,\dots, R$, 
  \begin{itemize}
  \item Simulate $z^{(r)}$ from the prior predictive distribution $p(z)$ for $z$.  
  \item Compute the posterior approximation $\widetilde{p}(\theta|z^{(r)})$ using (\ref{mixture-conditional}).
  \item Compute $\widetilde{G}^{(r)}=\widetilde{\text{KL}}(\widetilde{p}(\theta|z^{(r)})||\widetilde{p}(\theta))$ using 
  (\ref{KL-approx}).   
  \end{itemize}
  \item Compute $\widetilde{p}(\theta|z_{\text{obs}})$ using (\ref{mixture-conditional}),  
  $\widetilde{G}_{\text{obs}}=\widetilde{\text{KL}}(\widetilde{p}(\theta|z_{\text{obs}})|| \widetilde{p}(\theta))$ using (\ref{KL-approx}) and
  then
  \begin{align*}
  \widetilde{p}_{\text{KL}} & =\frac{1}{R}\sum_{r=1}^R I(\widetilde{G}^{(r)}\geq \widetilde{G}_{\text{obs}}).
  \end{align*}
  \end{enumerate}
  
\end{algorithm}

A similar approximate implementation of the conflict check based on prior-to-posterior divergences was considered
in \cite{nott+wee16}.  In that case, however, the likelihood was tractable and
the mixture posterior approximations were obtained by learning variational approximations independently for each simulated
prior predictive dataset in the Monte Carlo approximation of the tail probability (\ref{prior-to-posterior-tail}).  
Here our mixture approximations are obtained in quite a different way, and furthermore they are extremely fast
to compute for every new dataset once the mixture approximation to the joint distribution of $(\theta,z)$ has
been obtained.  This is important in the application we discuss next, which is searching for weakly informative
prior distributions, an application which was not considered in the work of \cite{nott+wee16}.

\section{Weakly informative priors}

\subsection{Weakly informative priors from prior-data conflict checks}

Weakly informative priors were first considered by \cite{gelman06}, conceived as prior distributions 
which put some prior information
into an analysis, but less than the analyst actually possesses.  \cite{evans+j11} gave a precise definition of a 
weakly informative prior with respect to a base prior used for an analysis in terms of prior-data conflict checks.  We discuss
this definition now.  

Let $p_B(\theta)$ denote the elicited informative prior (called the baseline prior) used in the analysis.  Let $p_W(\theta)$ denote
some alternative prior.  Suppose that $M$ is a minimal sufficient statistic.  Write $p_B(m)$ and $p_W(m)$ for the prior predictive
densities for $M$ for the priors $p_B(\theta)$ and $p_W(\theta)$ respectively.  \cite{evans+m06} consider using the prior
predictive density ordinate for $M$ as the statistic for a prior-data conflict check, and this is also used in the work
of \cite{evans+j11}.  So if the prior $p_j(\theta)$ is used for the analysis, $j=B,W$ then a tail probability for the prior-data
conflict check is computed as
\begin{align*}
 p_j & = P(p_j(M)\leq p_j(m_{\text{obs}})),\;\;\;\;M\sim p_j(m),
\end{align*}
where $m_{\text{obs}}$ is the observed value for $M$, as this determines whether or not 
$m_{\text{obs}}$ lies in a region with low probability with respect to $p_j(m)$.  
The definition of a weakly informative prior with respect to the base prior given in \cite{evans+j11} is based on the idea
that for data simulated under the base prior, there should be a reduction in the proportion of prior-data conflicts when
the data are analyzed under the alternative prior rather than the base prior.  

Suppose a conflict occurs if a $p$-value for a prior-data conflict check is less than $\alpha$ for some cutoff $\alpha$.  
Let $x_\alpha$ be the $\alpha$-quantile of the random variable
$P_B(M')$, $M'\sim p_B(m)$, where 
\begin{align*}
  P_B(M') & =P(p_B(M)\leq p_B(M'))\;\;\;\;M\sim p_B(m).  
\end{align*}
The distribution of $P_B(M')$ is that of the conflict $p$-value that is obtained
when $p_B(\theta)$ is used in the analysis, and the data are simulated under the prior predictive for $p_B(\theta)$.  
If $M$ is continuous then $P_B(M')$ will be uniform on $[0,1]$.  
Next, consider the random variable $P_W(M')$, $M'\sim P_B(M)$, where
\begin{align*}
  P_W(M') & = P(p_W(M)\leq p_W(M'))\;\;\;\;M\sim p_W(m).
\end{align*}
The distribution of $P_W(M')$ is that of a conflict $p$-value for data generated under
the prior predictive for $p_B(\theta)$, when the analysis is done using $p_W(\theta)$.  

We say the prior $p_W(\theta)$ is weakly informative with respect to $p_B(\theta)$ at level $\alpha$ if 
\begin{align*}
  P(P_W(M')\leq x_\alpha)<\alpha,
 \end{align*}
which says that prior-data conflicts happen less often when the data are analyzed using $p_W(\theta)$ rather than $p_B(\theta)$, but the data are generated under $p_B(\theta)$. Instead of choosing a fixed level $\alpha$ one can also consider
 other stronger notions of uniform weak informativity -- see \cite{evans+j11} for details.  \cite{evans+j11} define the
 degree of weak informativity of $p_W(\theta)$ relative to $p_B(\theta)$ at level $\alpha$ to be
 \begin{align}
  W_\alpha = 1 - \frac{P(P_W(M')\leq x_\alpha)}{x_\alpha},  \label{dofwi}
\end{align}
which is the proportion of prior-data conflicts avoided by using $p_W(\theta)$ as the prior for the analysis, 
with data generated
under $p_B(\theta)$.  The degree of weak informativity of one prior with respect to another defined by (\ref{dofwi}) 
can be compared for
different choices of the alternative prior, which do not need to belong to the same parametric family.

\subsection{Weakly informative priors based on conflict checks}

In the formulation of weakly informative priors used in \cite{evans+j11}, the prior-data conflict check 
based on the prior predictive density ordinate for a minimal sufficient statistic can be replaced by some other 
prior-data conflict check.  We consider this now for our prior-to-posterior divergence conflict checks.   
Let us consider a family of priors $p(\theta|\gamma)$ for searching for a weakly informative prior, 
where $\gamma$ is an expansion parameter.  We will assume the prior expansion
will be chosen so that $p(\theta)$ corresponds to a prior within this family for some value $\gamma^{(0)}$ 
so that $p(\theta)=p(\theta|\gamma^{(0)})$ say, although this is not essential.  Dealing with a baseline
prior that does not belong to the family $p(\theta|\gamma)$ does not involve
any alteration to the procedure we suggest below.  In the case where the baseline prior is elicited, it seems
natural that the family $p(\theta|\gamma)$ should be an expansion of the baseline prior, since we
want to retain some of the information in the original prior.  We can also consider choosing
a weakly informative prior from a family that is a union of two different parametric families.

Write $\widetilde{G}(z,\gamma)$ for the statistic $\widetilde{G}$ at (\ref{prior-to-posterior-approx}) when
the prior used for the analysis is $p(\theta|\gamma)$.  
We have previously discussed in Section 2.2 how to compute $\widetilde{G}(z,\gamma^{(0)})$ for abitrary 
observed summary statistics $z$ by fitting a mixture model to simulated data 
$x_i=(\theta_i,z_i)\sim p(\theta|\gamma_0)p(z|\theta)$, 
$i=1,\dots, n$.  We now wish to approximate $\widetilde{G}(z,\gamma)$ for both arbitrary $z$ and $\gamma$.  
We will accomplish this by expanding the original statistical model hierarchically to include $\gamma$ as a parameter, 
giving the model $p(\gamma)p(\theta|\gamma) p(z|\gamma)$, where $p(\gamma)$ is a pseudo-prior for 
$\gamma$.  We call $p(\gamma)$ a pseudo-prior, since we employ it for purely computational reasons to enable
us to approximate conditional posterior densities $p(\theta|z,\gamma)$.  Proceeding in a similar
way to Section 2.2, we can simulate data 
$$x_i=(\gamma_i,\theta_i,z_i)\sim p(\gamma)p(\theta|\gamma)p(z|\theta),$$
$i=1,\dots, n$, 
fit a Gaussian mixture model to these data, and then use the conditional distribution of $\theta$ given
$z,\gamma$ in the mixture as an estimated posterior distribution given $z,\gamma$ and hence
compute $\widetilde{G}(z,\gamma)$.  

The prior $p(\theta|\gamma)$ will be said to be weakly informative at level $\alpha$ relative to $p(\theta)$ 
for the approximate divergence check (\ref{prior-to-posterior-tail-approx}) if the random variable 
$P_\gamma(z')$, $z'\sim \int p(\theta)p(z|\theta) d\theta$, where
\begin{align*}
  P_{\gamma}(z') & = P(\widetilde{G}(z,\gamma) \geq \widetilde{G}(z',\gamma)),\;\;\;\; z\sim \int p(\theta|\gamma)p(z|\theta)\,d\theta,
\end{align*}
satisfies
\begin{align*}
  P(P_{\gamma}(z') \leq x_{\alpha}) <\alpha,
\end{align*}
where $x_{\alpha}$ is the $\alpha$-quantile of $P_{\gamma^{(0)}}$.   To approximate the distribution of 
$P_\gamma(z')$, we need to simulate values for $z'\sim \int p(\theta)p(z|\theta)\,d\theta$, and then for each
of these simulations we must approximate the $p$-value (\ref{prior-to-posterior-tail-approx}) using 
Algorithm \ref{approx-check} to get a Monte Carlo empirical distribution approximating the distribution of 
$P_\gamma(z')$.

The degree of weak informativity of $p(\theta|\gamma)$ at level $\alpha$ with respect to $p(\theta)=p(\theta|\gamma^{(0)})$ 
for the approximate divergence check (\ref{prior-to-posterior-tail-approx}) is, similar
to before, defined to be 
\begin{align*}
  W_{\alpha}(\gamma) & = 1-\frac{P(P_{\gamma}(z') \leq x_{\alpha})}{x_{\alpha}}.
\end{align*}
It seems reasonable to try to choose 
a prior $p(\theta|\gamma)$ weakly informative compared to $p(\theta)$ by choosing $\gamma$ such
that
\begin{align}
  & W_{\alpha}(\gamma)>\delta,  \label{wicriterion}
\end{align}
which would ensure that the proportion of conflicts is reduced by $\delta$ when data is simulated
under the base prior and the analysis is done under the alternative prior.  The constant
$\delta$ needs to be chosen and choosing $\delta=0.5$ would require reducing the proportion of conflicts by half, for example.
If it is not possible to find any prior satisfying (\ref{wicriterion}) we can look at maximizing 
$W_{\alpha}(\gamma)$.  
Later we consider checking the criterion (\ref{wicriterion}) at a finite number of candidate values for $\gamma$ chosen
as a maximin latin hypercube design covering some rectangular search region.  

\section{Examples}

\subsection{Logistic regression example}

We consider a logistic regression model as a first illustration of our methodology.  Although the likelihood
is tractable, we consider this example since weakly informative priors have been developed
for this model in the literature, and it is interesting to compare the priors obtained
using our approach with those in previous work.   We develop a weakly informative
prior in the context of a design from a real data set.  \cite{racine+gfs86} considered a 
bioassay experiment in which 5 animals at each of 4 dose levels were exposed to a toxin.  For the purposes
of considering weakly informative prior specification below we consider a hypothetical increase in the
number of animals at each dose to $20$.  This is to make the continuity assumption involved in a joint
modelling of data and parameters as a Gaussian mixture more reasonable.  At each dose, the number of 
deaths was recorded.  Writing $y_i$ for the number of deaths at dose $d_i$, the model is $y_i\sim \text{Binomial}(20,p_i)$, 
$\text{logit}(p_i)=\theta_1+\theta_2 d_i$, where the dose values have been log transformed, centred and scaled
similar to \cite{gelman+jps08}.

Consider a prior distribution for $\theta=(\theta_1,\theta_2)^\top$ of the form $p(\theta|\gamma)=p(\theta_1|\gamma_1)
p(\theta_2|\gamma_2)$ where $p(\theta_j|\gamma_j)=\phi(\theta_j;0,\gamma_j^2)$, $j=1,2$ 
with $\phi(x;\mu,\sigma^2)$ denoting
the normal density with mean $\mu$ and variance $\sigma^2$.  We use the base prior $\gamma^{(0)}=(1,1)$.  
Next consider a uniform grid of $50$ equally spaced values for $\gamma_1$ on the range $[0.5,10]$ and of $100$
equally spaced values for $\gamma_2$ on the range $[0.5,20]$.  From these
we can form a corresponding two-dimensional grid on $[0.5,10]\times [0.5,20]$.  
For each $\gamma$ on the two-dimensional grid, we estimate the degree of weak informativity of $p(\theta|\gamma)$ with
respect to the base prior at level $0.05$.  

Making the baseline variance parameters either larger or smaller can resolve a conflict in some instances.  
To get some intuition
for this, consider the simple case of a logistic regression without covariates, $\text{logit}(p_i)=\theta_1$, 
with a normal prior $N(0,\gamma_1^2)$  on $\theta_1$.  As $\gamma_1\rightarrow\infty$, most prior mass is on large values of $|\theta_1|$, which corresponds to probabilities close to zero or one.  On the other hand, choosing $\gamma_1\rightarrow 0$ 
gives a prior on the probability concentrated around $0.5$.  So we can see that choosing $\gamma_1$ either very large or close to
zero results in a highly informative prior, and so a choice of the prior variance parameter that avoids these extremes 
is necessary for a weakly informative choice.  See \cite{al-labadi+be18} for some related discussion.
For computing the approximate tail probabilities $\widetilde{p}_{\text{KL}}$ 
at (\ref{prior-to-posterior-tail-approx}), we used $R=1000$ prior predictive simulations.  
The mixture approximation to the joint distribution was trained using the R package \texttt{mclust} \citep{scrucca+fme16}
based on $100,000$ simulations from the model, and a uniform distribution on $[0.5,10]\times [0.5,20]$
was assumed for a pseudo-prior distribution for $\gamma$ in the mixture modelling.  The number of clusters 
was chosen using the default method implemented by the \texttt{mclustBIC} function in \texttt{mclust}, searching
up to a maximum of 15 clusters and considering 14 different possible choices for the mixture component covariance
structure.  In our later examples we use a similar approach to choosing the number of components.   The final model chose by BIC contained 14 mixture components here.  

Figure 1 plots $W_{0.05}(\gamma)$, for the mixture model chosen by BIC as well as a mixture model with 10 components 
to explore sensitivity of estimates of weak informativity to the number of mixture components used.  Little sensitivity is observed, 
particularly in the region where the degree of weak informativity is large,  if a
sufficiently large number of mixture components is chosen.  
Figures 2 and 4 in  \cite{nott+dme18} and \cite{evans+j11} respectively are qualitatively similar to Figure 1, although
the definition of a weakly informative prior depends on the  prior-data conflict check used, and
our check is different to that used by these authors.
From Figure 1 we see that making the variance parameters $\gamma_1$ and $\gamma_2$ somewhat
larger than their baseline values leads to a weakly informative prior.  However, if these parameters are made too
large this does not lead to a weakly informative prior, consistent with the intuition obtained
from the case discussed above of a logistic regression with an intercept only.  In situations where $\gamma$ is higher-dimensional,
it is not possible to evaluate the degree of weak informativity on a grid.  In these cases we generate a certain number of values
according to a minimax latin hypercube or some other space-filling design \citep{santner2003} 
to cover the search space for $\gamma$, and evaluate the degree of weak informativity on the design points.  
Generating 100 minimax latin hypercube design points in this example on $[0.5,10]\times [0.5,10]$ and choosing
the value for $\gamma$ maximizing the degree of weak informativity for the score checks with
respect to $\gamma_1$ and $\gamma_2$ gave a value $\gamma=(2.6,2.5)$.  

\begin{figure}
\centering
\begin{tabular}{cc}
\includegraphics[width=0.5\textwidth]{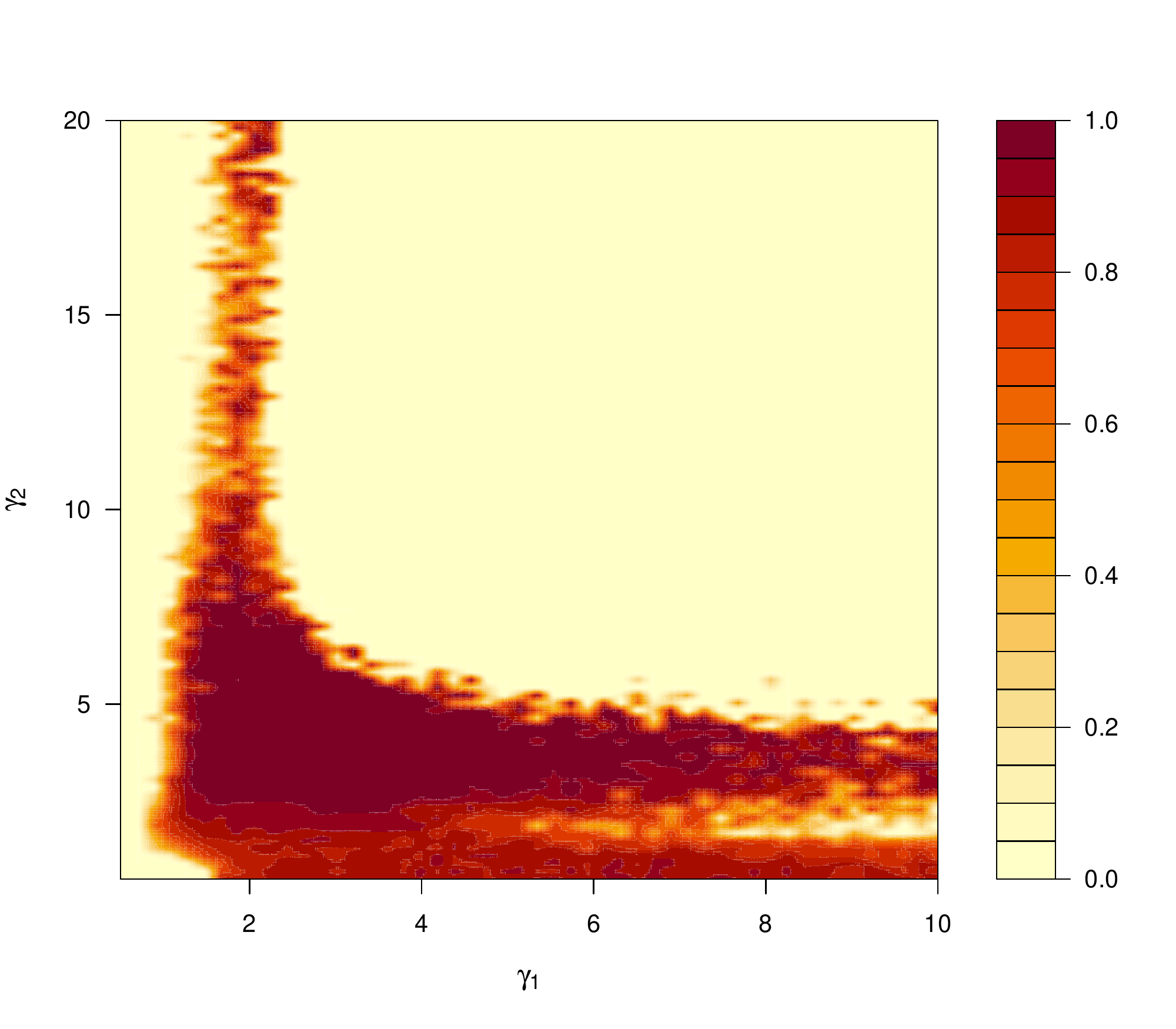} & \includegraphics[width=0.5\textwidth]{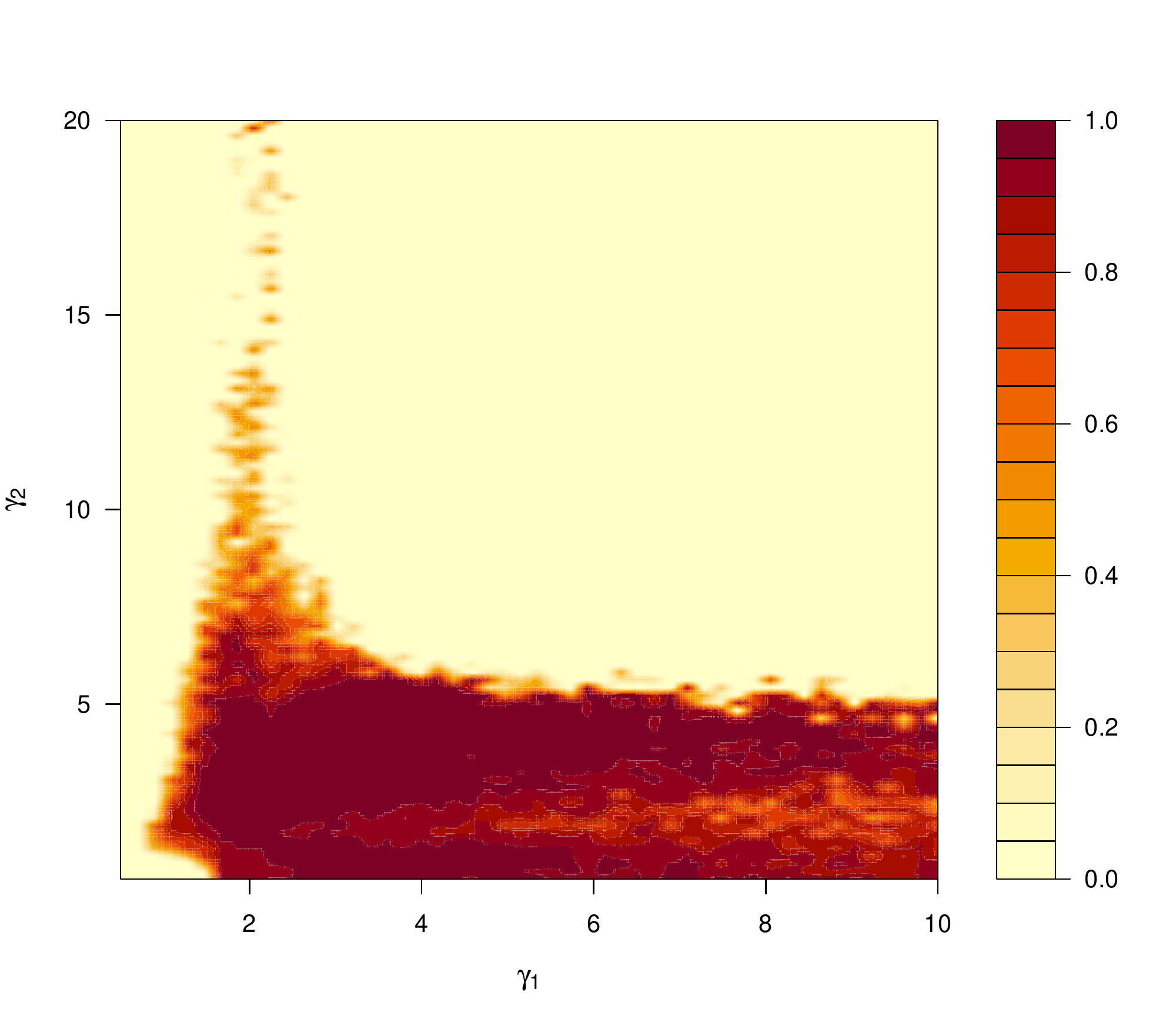}
\end{tabular}
\caption{\label{wi-prior-logistic}Degree of weak informativity for conflict check for logistic regression example with 14 
mixture components (left) and 10 mixture components (right).  The 14 component model was chosen by BIC.}
\end{figure}

\subsection{Multivariate $g$-and-$k$ example}

The $g$-and-$k$ distribution \citep{rayner+m02} is defined through its quantile function, 
$$Q(p;A,B,g,k)=A+B\left(1+c\frac{1-\exp(-g z(p))}{1+\exp(-gz(p))}\right)(1+z(p)^2)^k z(p), \;\;\;p\in (0,1),$$
where $z(p)=\Phi^{-1}(p)$ with $\Phi(\cdot)$ the standard normal distributon function, 
and $A$, $B$, $g$ and $k$ are location, scale, skewness and kurtosis
parameters, with $B>0$.  The constant $c$ is conventionally fixed at $0.8$, which results in the constraint $k>-0.5$.  
The closed form quantile function makes simulation from the distribution easy using the inversion method by computing
$Q(U;A,B,g,k)$ for $U\sim U[0,1]$.  This makes likelihood-free inference methods
attractive \citep{allingham+km09}.   Although it is possible to calculate the density function numerically with
sufficient computational effort
\citep{prangle17}, an additional motivation for using likelihood-free methods in this example is to robustify a 
Bayesian analysis to outliers.  The octile-based summary statistics described below allow a robust Bayesian
analysis where inference is insensitive to extreme outliers, and the summary statistic likelihood is intractable, leading
to an interest in likelihood-free inference methods.

We consider here the multivariate $g$-and-$k$ model described in \cite{drovandi+p11}.  Their model
uses a univarviate $g$-and-$k$ distribution for each marginal, and a Gaussian copula with a correlation matrix
$C$ for the dependence structure.  Precisely, let $y_i$, $i=1,\dots, n$, be the data, where $y_i=(y_{i1},\dots, y_{iJ})^\top$.  
The values $y_{ij}$, $i=1,\dots, n$, are iid and follow a univariate $g$-and-$k$ distribution with 
parameters $\theta_j=(A_j,B_j,g_j,k_j)$.  We
write the density of $y_{ij}$ as $f(y_{ij};\theta_j)$, with corresponding distribution function
$F(y_{ij};\theta_j)$.  
Define $\theta=(\theta_1^\top,\dots, \theta_J^\top,C)$, 
and then the joint density of $y_i$ is
$$f(y_i;\theta)=|C|^{-1/2} \exp\left( -\frac{1}{2} \eta_i^\top (I-C^{-1})\eta_i \right) \prod_{j=1}^J  f(y_{ij};\theta_j),$$
where $\eta_i=(\eta_{i1},\dots, \eta_{iJ})^\top$, with $\eta_{ij}=\Phi^{-1}(F(y_{ij};\theta_j))$.  
If $Z=(Z_1,\dots, Z_J)\sim N(0,C)$, and we compute
compute $(F^{-1}(\Phi(Z_1);\theta_1),\dots, F^{-1}(\Phi(Z_J);\theta_J)^\top$, then this produces
a simulation from the model.

For a multivariate dataset of exchange rate returns discussed in \cite{drovandi+p11}, \cite{li+nfs15}
 consider prior densities for the $\theta_j$ that are independent for $j=1,\dots, J$, 
with $\theta_j$ uniform on $[-0.1,0.1]\times [0,0.05]\times [-1,1]\times [-0.2,0.5]$.  For the copula correlation matrix $C$, we
follow \cite{ong+ntsd16} and consider a normal prior on a spherical 
parametrization of the elements of $C$ \citep{pinheiro+b96} to make
the parameters unconstrained.  This is explained further below.  
We will consider a multivariate model with $J=3$ components, 
and the unconstrained parameters for this model will be denoted by $w=(w_1,w_2,w_3)$.  In a spherical
parametrization the parameters $w$ 
determine the correlation matrix $C$ through its lower-triangular Cholesky factor $L$, $C=LL^\top$, by 
$$L=\left[ \begin{array}{ccc}
  1 & 0 & 0 \\
  \cos \gamma_1 & \sin \gamma_1 & 0 \\
  \cos \gamma_2 & \sin\gamma_2 \cos\gamma_3 & \sin\gamma_2 \sin \gamma_3 \end{array}\right],$$
  where $\gamma_j=\pi/(1+\exp(-w_j))$, $j=1,2,3$. 
\cite{ong+ntsd16} considered a prior on $w$ which is multivariate normal, $N(0,(1.75)^2 I_3)$, where $I_q$ denotes the 
identity matrix of dimension $q$.  Although a uniform prior on the correlation matrix could be considered, when
$J$ is large it is preferable in many applications to use a prior that shrinks towards independence.

The transformation to make the parametrization of the correlation matrix
unconstrained makes valid prior specification easy in the mathematical sense.  However, 
the transformed parameters are not easy to relate to prior knowledge we would typically have, 
regarding the correlation parameters directly.  
This increases the possibility of specifying a prior distribution that is informative in ways that are not intended.
For a base prior in this example we will consider a multivariate normal distribution $N(0,(0.5)^2 I_3)$, which is more informative
than the prior used in \cite{ong+ntsd16}, and then search for a weakly informative prior relative to this base prior.
In searching for a weakly informative prior, we consider prior distributions of the form $N(0,\gamma^2 I_3)$, 
where the parameter $\gamma$ lies in the range $[0.5,5]$.  For summary statistics, we use the same
summary statistics as in \cite{ong+ntsd16}.  These are robust estimates of location, scale, skewness and kurtosis
based on octiles considered in \cite{drovandi+p11} for each marginal (4 summary statisics for each component), 
and rank correlations for all pairs of components (3 summary statistics).   There are 15 summary statistics in total.
Since we are interested in weakly informative priors for the correlation parameters, we consider conflict checks
based on the prior-to-posterior divergence for $w$, and we assume that all the information in the summary
statistics about $w$ is contained in the 3 rank correlation summary statistics summarizing the dependence
structure.  For approximating our Kullback-Leibler divergence statistics it is then only necessary to
consider approximating the joint distribution of $(\gamma,w,S(w))$, where we assume a 
pseudo-prior for $\gamma$ that is uniform on $[0.5,5]$ and $S(w)$ denotes the three-dimensional
vector of the pairwise rank correlations.  We use $100,000$ simulations of $\gamma$, $w$ and $S(w)$ from
the model to train the mixture model, and for approximating tail probabilities $\widetilde{p}_{\text{KL}}$ 
at (\ref{prior-to-posterior-tail-approx}), we used $R=1000$ prior predictive simulations.  

\begin{figure}
\centering
\begin{tabular}{c}
\includegraphics[width=0.6\textwidth]{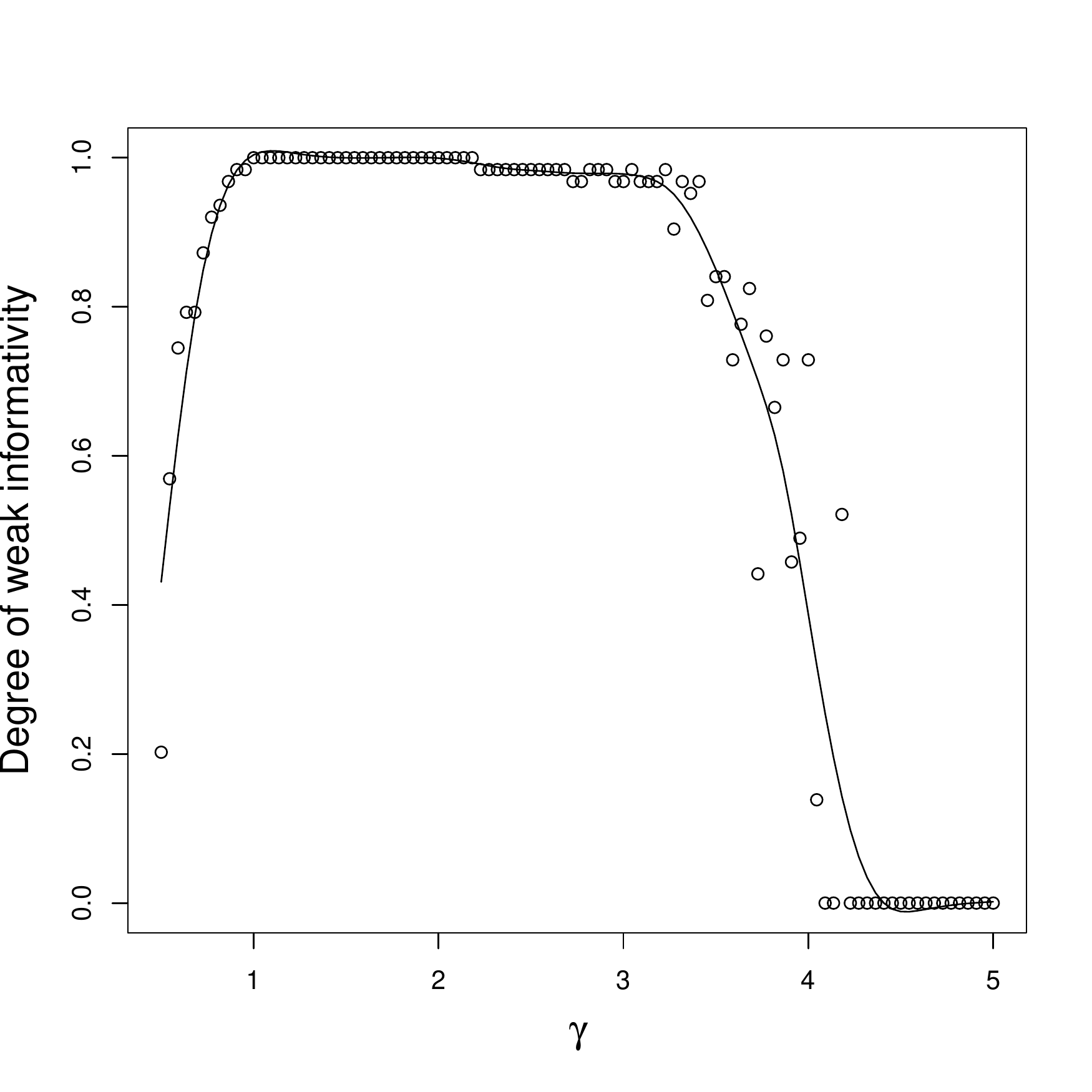}
\end{tabular}
\caption{\label{wi-prior-gnk}Degree of weak informativity for conflict check for multivariate $g$-and-$k$ example.}
\end{figure}

\begin{figure}
\centering
\begin{tabular}{c}
\includegraphics[width=0.7\textwidth]{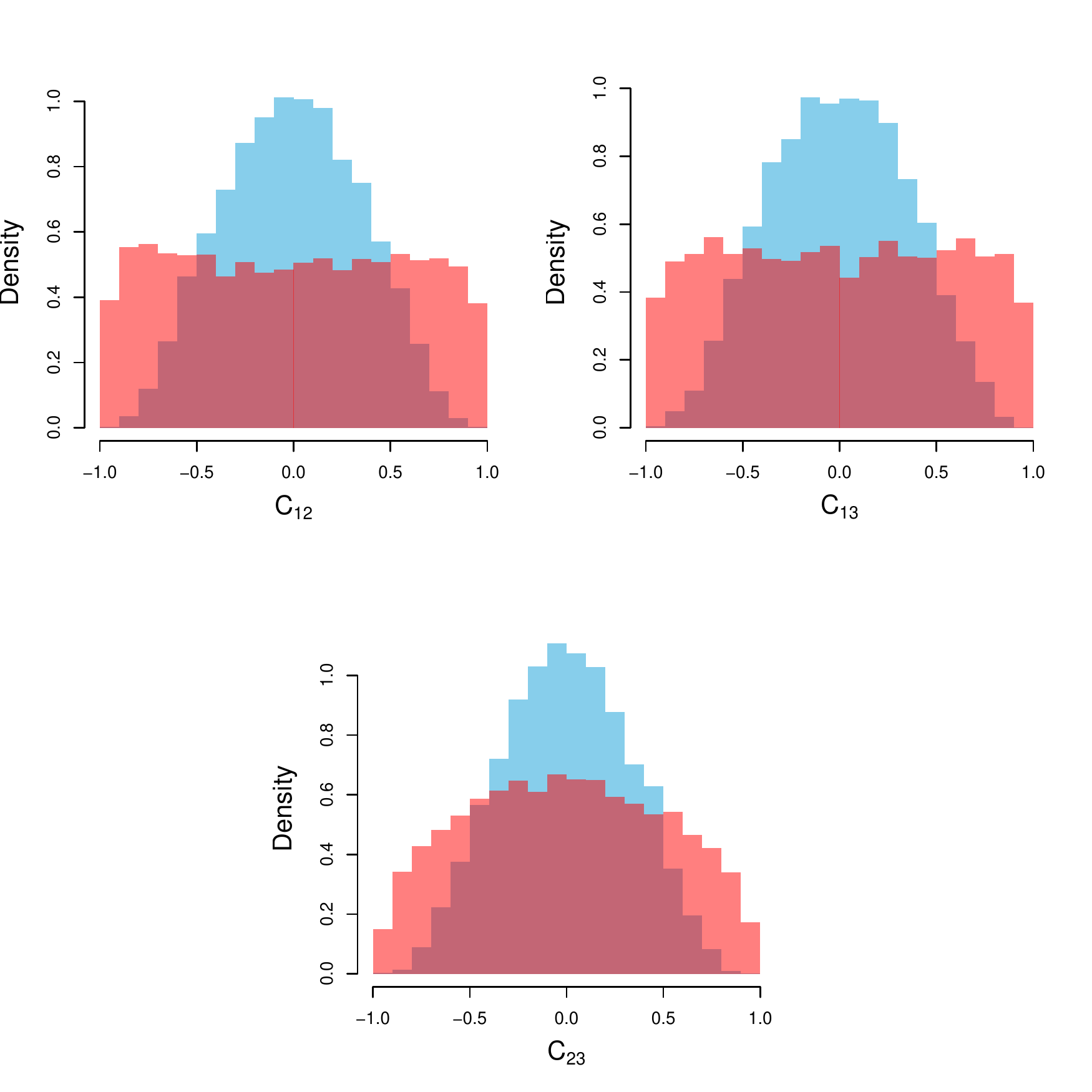}
\end{tabular}
\caption{\label{gnk-priors}Prior distribution on correlations for original ($\gamma=0.5$, blue) and weakly informative prior ($\gamma=1$, red) for multivariate $g$-and-$k$ example.}
\end{figure}

Figure \ref{wi-prior-gnk} plots the degree of weak informativity of the prior for different $\gamma$ with respect to the base
prior with $\gamma=0.5$.  Values of $\gamma$ in the range $1$ to $2$ here are maximally weakly informative
with respect to the base prior.  
For the base prior and a weakly informative prior with $\gamma=1$, we simulated $1000$ draws, and transformed these draws
to the corresponding correlations $C_{12}$, $C_{13}$ and $C_{23}$.  The result is shown in Figure \ref{gnk-priors}.
For the weakly informative prior, the implied marginal priors on the correlations are closer to uniform.  However, it is clear that the 
marginal prior distribution on the correlations depends on the ordering of the components, due to the way that the unconstrained
parameters are defined using a Cholesky decomposition.  

\subsection{Simple recruitment, boom and bust model}
 
\cite{fasiolo+pw16} discusses the motivation
for likelihood-free inference methods as an alternative to state space methods for likelihood estimation 
in time series models with complex nonlinear dynamics and chaotic behaviour, with likelihood-free methods sometimes
being preferable when there is low process noise or model misspecification.  
Our next example considers an ecological 
time series model representing the fluctuation of the population size of a certain group over time, 
considered in \cite{fasiolo+whb18} and \cite{an+nd18}, who both find that more flexible methods
than the synthetic likelihood method of \cite{wood10} and able to deal with non-Gaussian
distributions of summary statistics are needed.  

 Let $N_t$, $t=0,1, \dots$ represent population sizes
at discrete integer times $t$.   
Given $N_t$ and the parameters $\theta = (r, \kappa, \alpha, \beta)$, the 
conditional distribution of $N_{t+1}$ is
 \[   
N_{t+1} \sim 
     \begin{cases}
      \text{Poisson}(N_t(1+r))+\epsilon_t & \quad \text{if } N_t \leq \kappa\\
      \text{Binom}(N_t, \alpha)+\epsilon_t & \quad \text{if } N_t > \kappa\\
     \end{cases}
\]
where $\epsilon_t\sim \text{Poisson }(\beta).$  In this model $r$ is a growth parameter, $\kappa$ is a threshold where
exceedance of the threshold leads to a crash, $\alpha$ is a survival probability controlling the speed of the crash
and $\beta$ is the mean for a recruitment process.  
We consider a time series of length $250$, and in simulating from the model we use $50$ burn-in values 
after initializing the process at the integer part of the threshold $\kappa$.   

\cite{an+nd18} considered a prior 
uniform on $[0,1]\times [10,80]\times [0,1]\times [0,1]$.  
We change the $U[0,1]$ prior for $r$ to a $\text{Beta}(5,5)$ prior to obtain
the base prior for constructing a weakly informative alternative. 
The summary statistics $z$ are constructed following \cite{an+nd18}.   For a time series $x$ of length $T$, 
define differences and ratios $d_x =\{ x_i- x_{i-1}; i = 2, \dots, T\}$ and $r_x = \{x_i/x_{i-1}; i = 2, \dots,T\}$, 
respectively. We use the sample mean, variance, skewness and kurtosis of $x$, $d_x$ and $r_x$ as the summary statistics, 
so that $z$ is $12$-dimensional. 
 To search for a weakly informative prior, consider prior
distributions for $r$ of the form $r\sim \text{Beta}(\gamma,\gamma)$, 
so that the mean is fixed at $0.5$ but the variance changes with
$\gamma$.  

\begin{figure}
    \centering
    \includegraphics[width=12cm]{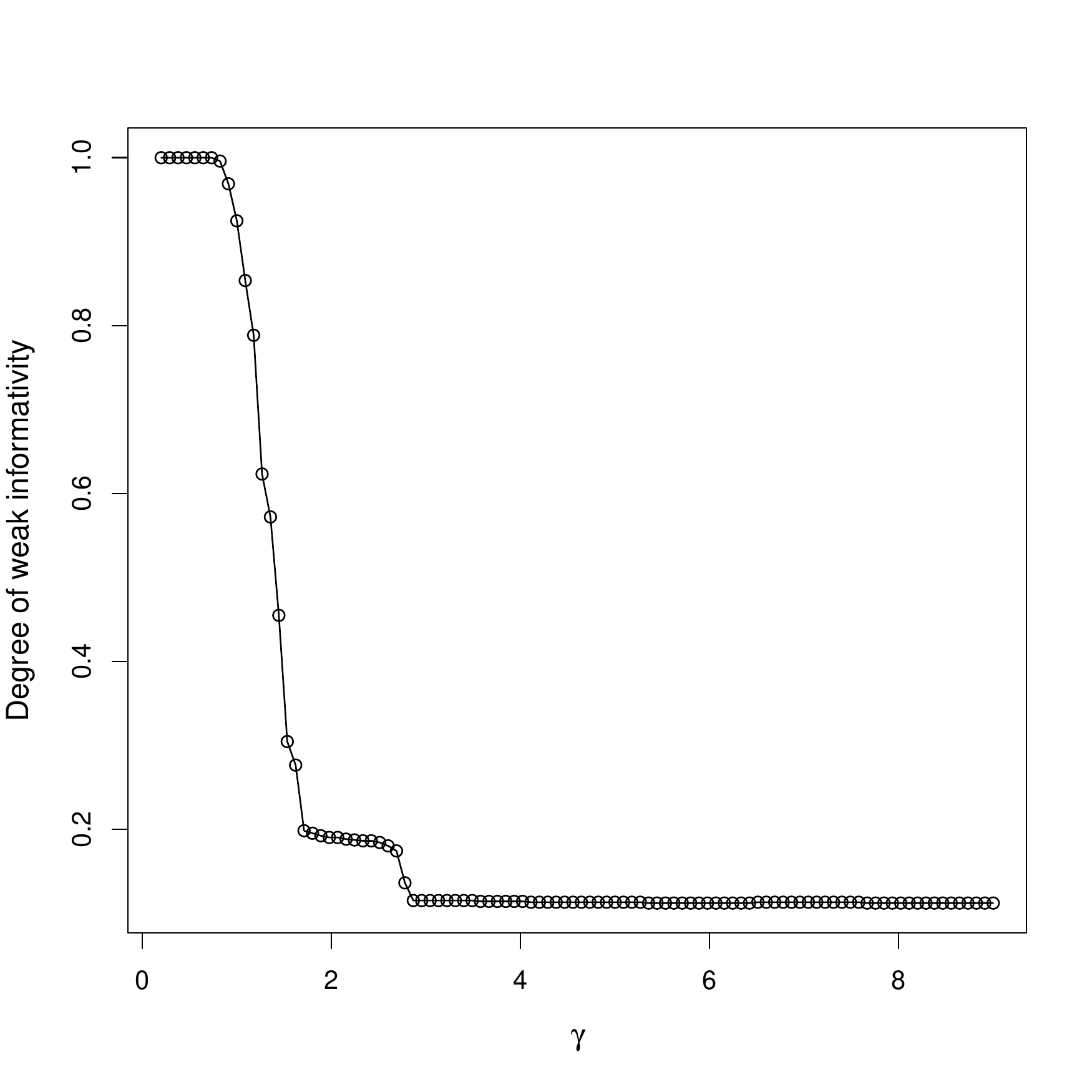}
    \caption{Degree of weak informativity for conflict check for boom and bust example.}
    \label{fig:degr}
\end{figure}

We use 100,000 simulations from the joint distribution of $\gamma, r$ and $z$ to train the mixture model, where 
a pseudo-prior uniform on $[0.2,9]$ was considered for $\gamma$.   For
approximating tail probabilities we used $R = 1000$ prior predictive simulations.  
 Figure \ref{fig:degr} plots the degree of weak informativity of the prior for different $\gamma$ with respect
to the base prior with $\gamma =5.$ 

\begin{figure}
    \centering
    \includegraphics[width=12cm]{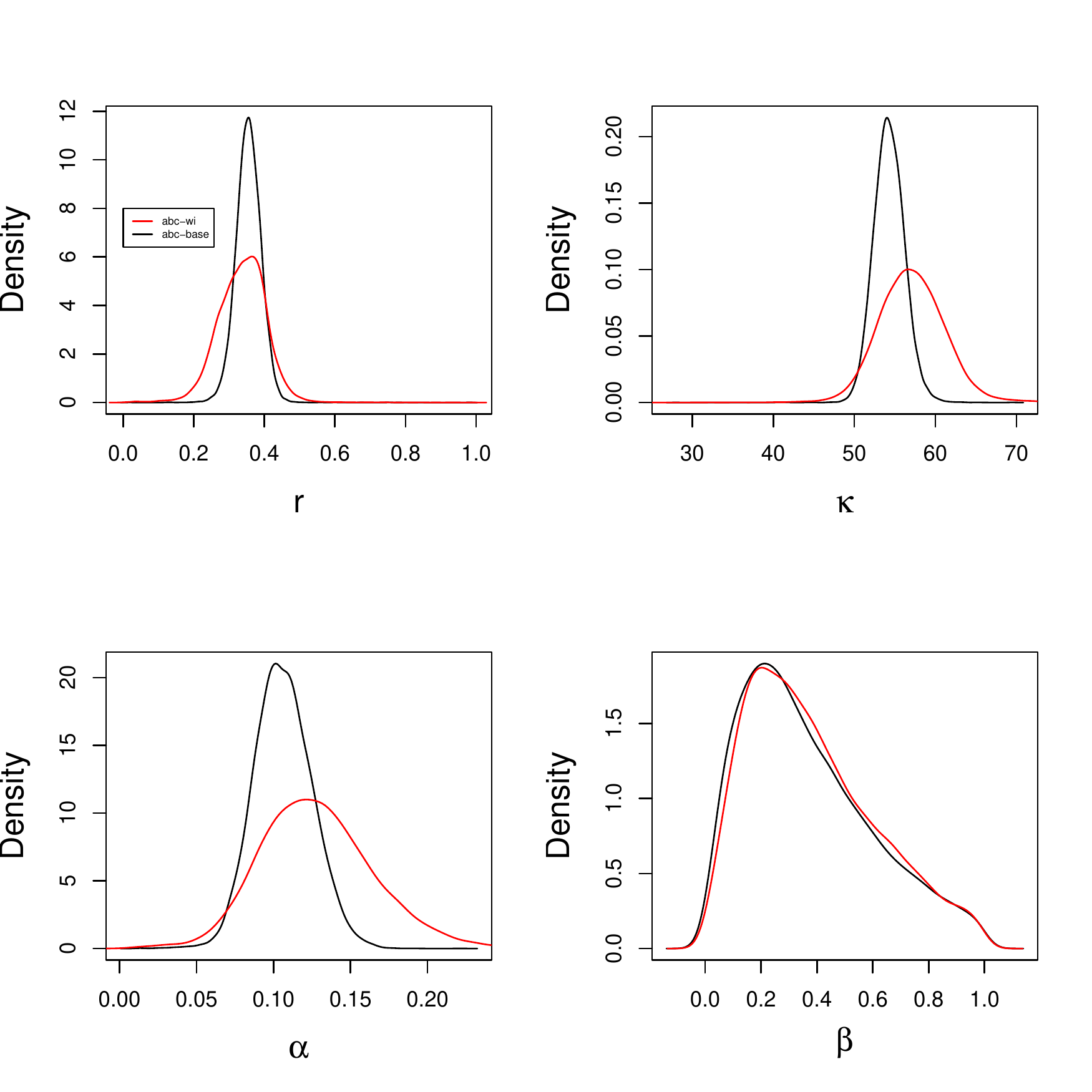}
    \caption{Estimated univariate posterior marginal densities for boom and bust example.}
    \label{fig:uniabc}
\end{figure}

\begin{figure}
    \centering
    \includegraphics[width=12cm]{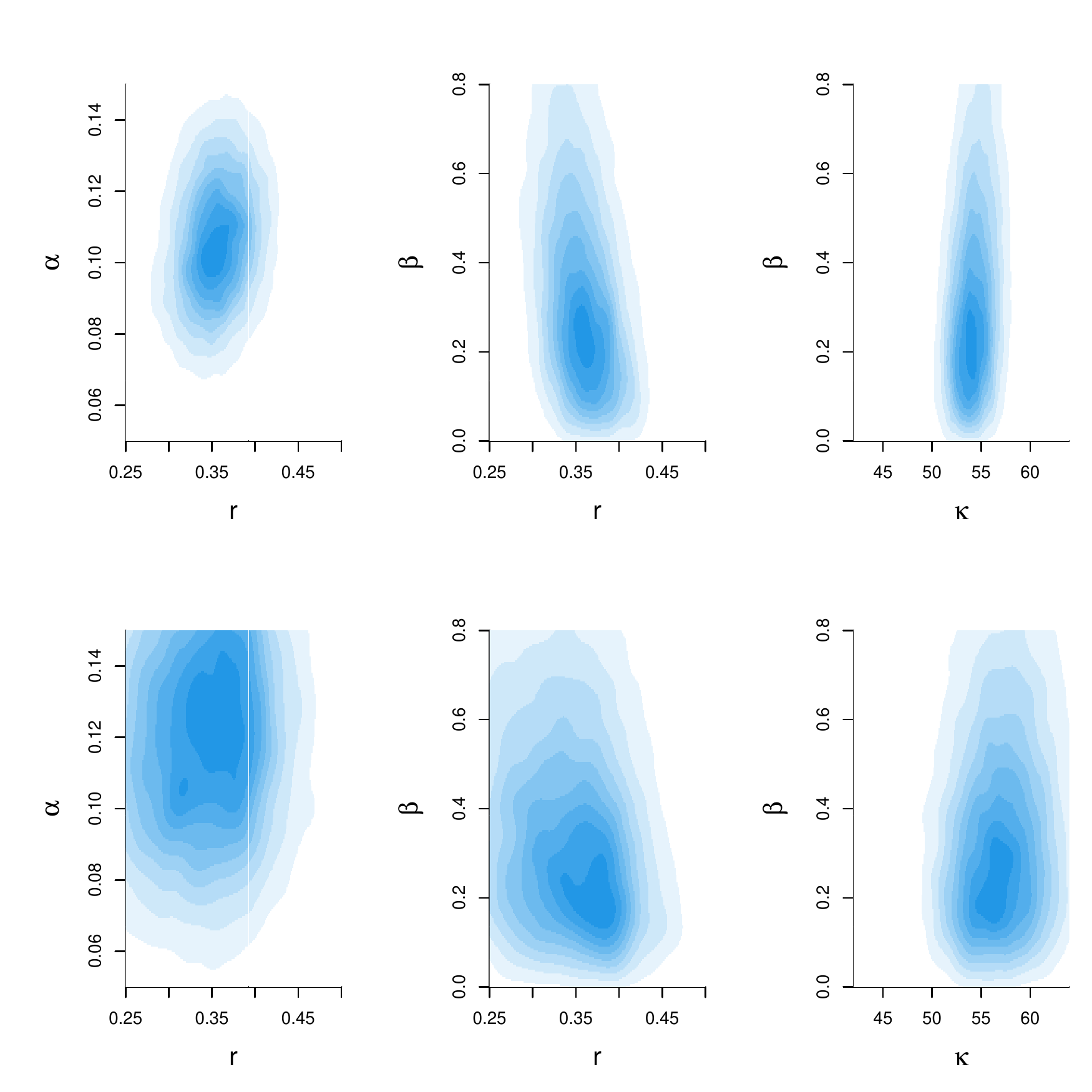}
    \caption{Estimated bivariate posterior marginal densities for boom and bust example. 
    The top and bottom rows shows estimates for the baseline and weakly informative priors 
    respectively.}
    \label{fig:bivabc}
\end{figure}

We choose here a value of $\gamma =0.2$ as a weakly informative choice.   
To show that using a weakly informative prior can make a difference for Bayesian inference, 
Figure \ref{fig:uniabc} shows, for a simulated time series, the 
estimated univariate posterior densities for the two prior distributions, while Figure
\ref{fig:bivabc} shows estimated bivariate posterior densities.
The simulated time series is of length $250$ with true parameter values 
$r = 0.4, \kappa = 50, \alpha = 0.09$ and $\beta = 0.05$
and the posterior density estimation was done using an ABC method.
The ABC analysis was based on 500,000 samples from the prior and a neural network
regression adjustment using the {\tt abc} function
in the {\tt abc} R pacakge \citep{csillery+fb12} with a tolerance of $0.05$ and other algorithmic settings
at default values.  
Given the complex interactions between the parameters, changing the marginal
prior on $r$ affects posterior inference not just for $r$ but also for the other parameters, particularly
$\kappa$ and $\alpha$.

\smallskip

\smallskip

\section{Discussion}

Informative priors are often needed in typical applications of likelihood-free inference.  The complex models for which
likelihood-free inference methods are useful often contain weakly identified parameters where the regularization
provided by an informative prior is valuable.  Some likelihood-free algorithms require a proper prior, and the computational
efficiency of such algorithms may depend on how informative the prior is, which creates the temptation to specify
priors for computational convenience.  It seems important then to develop new tools for assessing the sensitivity
of Bayesian inferences to the prior in the likelihood-free setting.  We have developed here methods for checking
for prior-data conflict, as well as methods for specifying weakly informative priors relative to the prior used in the analysis
which are useful for sensitivity analyses and for revealing possible deficiencies in prior elicitation and model understanding. 

Our approach to making the computations tractable in our conflict checks and in searching for weakly informative
priors uses Gaussian mixture approximations to posterior distributions and this may be rather crude, particularly with 
high-dimenisonal parameters or summary statistics.  While rough calculations may be good enough for
diagnostics and exploring alternative prior specifications, an interesting direction for future work is to investigate
better approaches to the likelihood-free inference while still allowing the repeated calculation of posterior densities
for different data that is necessary here.  

\subsection*{Acknowledgements}

We thank Anne Presanis for her comments on an earlier version of this manuscript.  


\bibliographystyle{chicago}
\bibliography{conflict-abc}

\begin{thebibliography}{}

\bibitem[\protect\citeauthoryear{Al-Labadi, Baskurt, and Evans}{Al-Labadi
  et~al.}{2018}]{al-labadi+be18}
Al-Labadi, L., Z.~Baskurt, and M.~Evans (2018).
\newblock Statistical reasoning: Choosing and checking the ingredients,
  inferences based on a measure of statistical evidence with some applications.
\newblock {\em Entropy\/}~{\em 20\/}(4), 1--19.

\bibitem[\protect\citeauthoryear{Al~Labadi and Evans}{Al~Labadi and
  Evans}{2017}]{allabadi+e17}
Al~Labadi, L. and M.~Evans (2017).
\newblock Optimal robustness results for relative belief inferences and the
  relationship to prior-data conflict.
\newblock {\em Bayesian Analysis\/}~{\em 12\/}(3), 705--728.

\bibitem[\protect\citeauthoryear{Allingham, King, and Mengersen}{Allingham
  et~al.}{2009}]{allingham+km09}
Allingham, D.~R., A.~R. King, and K.~L. Mengersen (2009).
\newblock Bayesian estimation of quantile distributions.
\newblock {\em Statistics and Computing\/}~{\em 19}, 189--201.

\bibitem[\protect\citeauthoryear{An, Nott, and Drovandi}{An
  et~al.}{2020}]{an+nd18}
An, Z., D.~J. Nott, and C.~Drovandi (2020).
\newblock Robust {B}ayesian synthetic likelihood via a semi-parametric
  approach.
\newblock {\em Statistics and Computing\/}~{\em 30}, 543--557.

\bibitem[\protect\citeauthoryear{Bayarri and Castellanos}{Bayarri and
  Castellanos}{2007}]{bayarri+c07}
Bayarri, M.~J. and M.~E. Castellanos (2007).
\newblock Bayesian checking of the second levels of hierarchical models.
\newblock {\em Statistical Science\/}~{\em 22}, 322--343.

\bibitem[\protect\citeauthoryear{Beaumont, Zhang, and Balding}{Beaumont
  et~al.}{2002}]{beaumont+zb02}
Beaumont, M.~A., W.~Zhang, and D.~J. Balding (2002).
\newblock Approximate {Bayesian} computation in population genetics.
\newblock {\em Genetics\/}~{\em 162}, 2025--2035.

\bibitem[\protect\citeauthoryear{Berger, Bernardo, and Sun}{Berger
  et~al.}{2009}]{berger+bs09}
Berger, J.~O., J.~M. Bernardo, and D.~Sun (2009).
\newblock The formal definition of reference priors.
\newblock {\em The Annals of Statistics\/}~{\em 37\/}(2), 905--938.

\bibitem[\protect\citeauthoryear{Blum, Nunes, Prangle, and Sisson}{Blum
  et~al.}{2013}]{blum+nps13}
Blum, M. G.~B., M.~A. Nunes, D.~Prangle, and S.~A. Sisson (2013).
\newblock A comparative review of dimension reduction methods in approximate
  {B}ayesian computation.
\newblock {\em Statistical Science\/}~{\em 28\/}(2), 189--208.

\bibitem[\protect\citeauthoryear{Bonassi and West}{Bonassi and
  West}{2015}]{bonassi+w15}
Bonassi, F.~V. and M.~West (2015).
\newblock Sequential {M}onte {C}arlo with adaptive weights for approximate
  {B}ayesian computation.
\newblock {\em Bayesian Analysis\/}~{\em 10\/}(1), 171 -- 187.

\bibitem[\protect\citeauthoryear{Bonassi, You, and West}{Bonassi
  et~al.}{2011}]{bonassi+yw11}
Bonassi, F.~V., L.~You, and M.~West (2011).
\newblock Bayesian learning from marginal data in bionetwork models.
\newblock {\em Statistical Applications in Genetics and Molecular
  Biology\/}~{\em 10\/}(1).

\bibitem[\protect\citeauthoryear{Box}{Box}{1980}]{box80}
Box, G. E.~P. (1980).
\newblock {Sampling and Bayes' inference in scientific modelling and robustness
  (with discussion)}.
\newblock {\em Journal of the Royal Statistical Society, Series A\/}~{\em 143},
  383--430.

\bibitem[\protect\citeauthoryear{Clarke and Gustafson}{Clarke and
  Gustafson}{1998}]{clarke+g98}
Clarke, B. and P.~Gustafson (1998).
\newblock On the overall sensitivity of the posterior distribution to its
  inputs.
\newblock {\em Journal of Statistical Planning and Inference\/}~{\em 71},
  137--150.

\bibitem[\protect\citeauthoryear{Csill{\'e}ry, Fran{\c{c}}ois, and
  Blum}{Csill{\'e}ry et~al.}{2012}]{csillery+fb12}
Csill{\'e}ry, K., O.~Fran{\c{c}}ois, and M.~G.~B. Blum (2012).
\newblock {ABC}: an {R} package for approximate {B}ayesian computation ({ABC}).
\newblock {\em Methods in Ecology and Evolution\/}~{\em 3}, 475--479.

\bibitem[\protect\citeauthoryear{Drovandi and Pettitt}{Drovandi and
  Pettitt}{2011}]{drovandi+p11}
Drovandi, C.~C. and A.~N. Pettitt (2011).
\newblock Likelihood-free {Bayesian} estimation of multivariate quantile
  distributions.
\newblock {\em Computational Statistics and Data Analysis\/}~{\em 55},
  2541--2556.

\bibitem[\protect\citeauthoryear{Evans}{Evans}{2015}]{evans15}
Evans, M. (2015).
\newblock {\em Measuring Statistical Evidence Using Relative Belief}.
\newblock Taylor \& Francis.

\bibitem[\protect\citeauthoryear{Evans and Jang}{Evans and
  Jang}{2010}]{evans+j10}
Evans, M. and G.~H. Jang (2010).
\newblock Invariant p-values for model checking.
\newblock {\em The Annals of Statistics\/}~{\em 38}, 512--525.

\bibitem[\protect\citeauthoryear{Evans and Jang}{Evans and
  Jang}{2011}]{evans+j11}
Evans, M. and G.~H. Jang (2011).
\newblock Weak informativity and the information in one prior relative to
  another.
\newblock {\em Statistical Science\/}~{\em 26}, 423--439.

\bibitem[\protect\citeauthoryear{Evans and Moshonov}{Evans and
  Moshonov}{2006}]{evans+m06}
Evans, M. and H.~Moshonov (2006).
\newblock Checking for prior-data conflict.
\newblock {\em Bayesian Analysis\/}~{\em 1}, 893--914.

\bibitem[\protect\citeauthoryear{Fan, Nott, and Sisson}{Fan
  et~al.}{2013}]{fan+ns13}
Fan, Y., D.~J. Nott, and S.~A. Sisson (2013).
\newblock Approximate {B}ayesian computation via regression density estimation.
\newblock {\em Stat\/}~{\em 2\/}(1), 34--48.

\bibitem[\protect\citeauthoryear{Fasiolo, Pya, and Wood}{Fasiolo
  et~al.}{2016}]{fasiolo+pw16}
Fasiolo, M., N.~Pya, and S.~N. Wood (2016).
\newblock A comparison of inferential methods for highly nonlinear state space
  models in ecology and epidemiology.
\newblock {\em Statistical Science\/}~{\em 31}, 96--118.

\bibitem[\protect\citeauthoryear{Fasiolo, Wood, Hartig, and Bravington}{Fasiolo
  et~al.}{2018}]{fasiolo+whb18}
Fasiolo, M., S.~N. Wood, F.~Hartig, and M.~V. Bravington (2018).
\newblock {An extended empirical saddlepoint approximation for intractable
  likelihoods}.
\newblock {\em Electronic Journal of Statistics\/}~{\em 12\/}(1), 1544 -- 1578.

\bibitem[\protect\citeauthoryear{Forbes, Nguyen, Nguyen, and Arbel}{Forbes
  et~al.}{2021}]{forbes+nna21}
Forbes, F., H.~D. Nguyen, T.~T. Nguyen, and J.~Arbel (2021).
\newblock {Approximate Bayesian computation with surrogate posteriors}.
\newblock Inria technical report, hal-03139256,
  \url{https://hal.archives-ouvertes.fr/hal-03139256v2/file/Gllim-ABC_v2_4HALApril2021.pdf}.

\bibitem[\protect\citeauthoryear{Gelman}{Gelman}{2006}]{gelman06}
Gelman, A. (2006).
\newblock Prior distributions for variance parameters in hierarchical models.
\newblock {\em Bayesian Analysis\/}~{\em 1\/}(3), 1--19.

\bibitem[\protect\citeauthoryear{Gelman, Jakulin, Pittau, and Su}{Gelman
  et~al.}{2008}]{gelman+jps08}
Gelman, A., A.~Jakulin, M.~G. Pittau, and Y.-S. Su (2008).
\newblock A weakly informative default prior distribution for logistic and
  other regression models.
\newblock {\em The Annals of Applied Statistics\/}~{\em 2}, 1360--1383.

\bibitem[\protect\citeauthoryear{Gelman, Meng, and Stern}{Gelman
  et~al.}{1996}]{gelman+ms96}
Gelman, A., X.-L. Meng, and H.~Stern (1996).
\newblock Posterior predictive assessment of model fitness via realized
  discrepancies.
\newblock {\em Statistica Sinica\/}~{\em 6}, 733--807.

\bibitem[\protect\citeauthoryear{G\r{a}semyr and Natvig}{G\r{a}semyr and
  Natvig}{2009}]{gasemyr+n09}
G\r{a}semyr, J. and B.~Natvig (2009).
\newblock Extensions of a conflict measure of inconsistencies in {B}ayesian
  hierarchical models.
\newblock {\em Scandinavian Journal of Statistics\/}~{\em 36}, 822--838.

\bibitem[\protect\citeauthoryear{He, Huo, and Yang}{He et~al.}{2021}]{he+hy21}
He, Z., S.~Huo, and T.~Yang (2021).
\newblock {An adaptive mixture-population Monte Carlo method for
  likelihood-free inference}.
\newblock arXiv:2112.00420.

\bibitem[\protect\citeauthoryear{Hershey and Olsen}{Hershey and
  Olsen}{2007}]{hershey+o07}
Hershey, J.~R. and P.~A. Olsen (2007).
\newblock Approximating the {K}ullback {L}eibler divergence between {G}aussian
  mixture models.
\newblock In {\em 2007 IEEE International Conference on Acoustics, Speech and
  Signal Processing - ICASSP '07}, Volume~4, pp.\  IV--317--IV--320.

\bibitem[\protect\citeauthoryear{Lavine}{Lavine}{1991}]{lavine91}
Lavine, M. (1991).
\newblock Sensitivity in {B}ayesian statistics: The prior and the likelihood.
\newblock {\em Journal of the American Statistical Association\/}~{\em
  86\/}(414), 396--399.

\bibitem[\protect\citeauthoryear{Lewis, MacEachern, and Lee}{Lewis
  et~al.}{2021}]{lewis+ml21}
Lewis, J.~R., S.~N. MacEachern, and Y.~Lee (2021).
\newblock {Bayesian Restricted Likelihood Methods: Conditioning on Insufficient
  Statistics in Bayesian Regression}.
\newblock {\em Bayesian Analysis\/}~(To appear).

\bibitem[\protect\citeauthoryear{Li, Nott, Fan, and Sisson}{Li
  et~al.}{2017}]{li+nfs15}
Li, J., D.~J. Nott, Y.~Fan, and S.~A. Sisson (2017).
\newblock {Extending approximate Bayesian computation methods to high
  dimensions via Gaussian copula}.
\newblock {\em Computational Statistics and Data Analysis\/}~{\em 106}, 77--89.

\bibitem[\protect\citeauthoryear{Marshall and Spiegelhalter}{Marshall and
  Spiegelhalter}{2007}]{marshall+s07}
Marshall, E.~C. and D.~J. Spiegelhalter (2007).
\newblock Identifying outliers in {B}ayesian hierarchical models: a
  simulation-based approach.
\newblock {\em Bayesian Analysis\/}~{\em 2}, 409--444.

\bibitem[\protect\citeauthoryear{McCulloch}{McCulloch}{1989}]{mcculloch89}
McCulloch, R.~E. (1989).
\newblock Local model influence.
\newblock {\em Journal of the American Statistical Association\/}~{\em
  84\/}(406), 473--478.

\bibitem[\protect\citeauthoryear{Nott, Drovandi, Mengersen, and Evans}{Nott
  et~al.}{2018}]{nott+dme18}
Nott, D.~J., C.~C. Drovandi, K.~Mengersen, and M.~Evans (2018).
\newblock Approximation of {B}ayesian predictive p-values using regression
  {ABC}.
\newblock {\em Bayesian Analysis\/}~{\em 13\/}(1), 59--83.

\bibitem[\protect\citeauthoryear{Nott, Wang, Evans, and Englert}{Nott
  et~al.}{2020}]{nott+wee16}
Nott, D.~J., X.~Wang, M.~Evans, and B.-G. Englert (2020).
\newblock Checking for prior-data conflict using prior-to-posterior
  divergences.
\newblock {\em Statistical Science\/}~{\em 35\/}(2), 234--253.

\bibitem[\protect\citeauthoryear{O'Hagan}{O'Hagan}{2003}]{ohagan03}
O'Hagan, A. (2003).
\newblock {HSS} model criticism (with discussion).
\newblock In P.~J. Green, N.~L. Hjort, and S.~T. Richardson (Eds.), {\em Highly
  Structured Stochastic Systems}, pp.\  423--453. Oxford University Press.

\bibitem[\protect\citeauthoryear{Ong, Nott, Tran, Sisson, and Drovandi}{Ong
  et~al.}{2018}]{ong+ntsd16}
Ong, V. M.-H., D.~J. Nott, M.-N. Tran, S.~Sisson, and C.~Drovandi (2018).
\newblock Variational {B}ayes with synthetic likelihood.
\newblock {\em Statistics and Computing\/}~{\em 28\/}(4), 971--988.

\bibitem[\protect\citeauthoryear{Pinheiro and Bates}{Pinheiro and
  Bates}{1996}]{pinheiro+b96}
Pinheiro, J.~C. and D.~M. Bates (1996).
\newblock Unconstrained parametrizations for variance-covariance matrices.
\newblock {\em Statistics and Computing\/}~{\em 6\/}(3), 289--296.

\bibitem[\protect\citeauthoryear{Prangle}{Prangle}{2017}]{prangle17}
Prangle, D. (2017).
\newblock {gk}: An {R} package for the g-and-k and generalised g-and-h
  distributions.
\newblock arXiv:1706.06889.

\bibitem[\protect\citeauthoryear{Prangle}{Prangle}{2018}]{prangle18}
Prangle, D. (2018).
\newblock Summary statistics in approximate {B}ayesian computation.
\newblock In S.~A. Sisson, Y.~Fan, and M.~A. Beaumont (Eds.), {\em Handbook of
  Approximate Bayesian Computation}. Chapman {\&} Hall/CRC.

\bibitem[\protect\citeauthoryear{Presanis, Ohlssen, Spiegelhalter, and
  Angelis}{Presanis et~al.}{2013}]{presanis+osd13}
Presanis, A.~M., D.~Ohlssen, D.~J. Spiegelhalter, and D.~D. Angelis (2013).
\newblock Conflict diagnostics in directed acyclic graphs, with applications in
  {B}ayesian evidence synthesis.
\newblock {\em Statistical Science\/}~{\em 28}, 376--397.

\bibitem[\protect\citeauthoryear{Price, Drovandi, Lee, and Nott}{Price
  et~al.}{2018}]{price+dln16}
Price, L.~F., C.~C. Drovandi, A.~C. Lee, and D.~J. Nott (2018).
\newblock Bayesian synthetic likelihood.
\newblock {\em Journal of Computational and Graphical Statistics\/}~{\em
  27\/}(1), 1--11.

\bibitem[\protect\citeauthoryear{Pritchard, Seielstad, Perez-Lezaun, and
  Feldman}{Pritchard et~al.}{1999}]{pritchard+spf99}
Pritchard, J.~K., M.~T. Seielstad, A.~Perez-Lezaun, and M.~W. Feldman (1999).
\newblock Population growth of human {Y} chromosomes: {A} study of {Y}
  chromosome microsatellites.
\newblock {\em Molecular Biology and Evolution\/}~{\em 16}, 1791--1798.

\bibitem[\protect\citeauthoryear{Racine, Grieve, Fl\"{u}hler, and Smith}{Racine
  et~al.}{1986}]{racine+gfs86}
Racine, A., A.~P. Grieve, H.~Fl\"{u}hler, and A.~F.~M. Smith (1986).
\newblock Bayesian methods in practice: Experiences in the pharmaceutical
  industry.
\newblock {\em Journal of the Royal Statistical Society, Series C\/}~{\em 35},
  93--150.

\bibitem[\protect\citeauthoryear{Rayner and MacGillivray}{Rayner and
  MacGillivray}{2002}]{rayner+m02}
Rayner, G. and H.~MacGillivray (2002).
\newblock Weighted quantile-based estimation for a class of transformation
  distributions.
\newblock {\em Computational Statistics \& Data Analysis\/}~{\em 39\/}(4),
  401--433.

\bibitem[\protect\citeauthoryear{Roos, Martins, Held, and Rue}{Roos
  et~al.}{2015}]{roos+mhr15}
Roos, M., T.~G. Martins, L.~Held, and H.~Rue (2015).
\newblock Sensitivity analysis for {B}ayesian hierarchical models.
\newblock {\em Bayesian Analysis\/}~{\em 10}, 321--349.

\bibitem[\protect\citeauthoryear{Santner, Williams, Notz, and Williams}{Santner
  et~al.}{2003}]{santner2003}
Santner, T.~J., B.~J. Williams, W.~I. Notz, and B.~J. Williams (2003).
\newblock {\em The design and analysis of computer experiments}.
\newblock Springer.

\bibitem[\protect\citeauthoryear{Scrucca, Fop, Murphy, and Raftery}{Scrucca
  et~al.}{2016}]{scrucca+fme16}
Scrucca, L., M.~Fop, T.~B. Murphy, and A.~E. Raftery (2016).
\newblock {mclust} 5: clustering, classification and density estimation using
  {G}aussian finite mixture models.
\newblock {\em The {R} Journal\/}~{\em 8\/}(1), 289--317.

\bibitem[\protect\citeauthoryear{Sisson, Fan, and Beaumont}{Sisson
  et~al.}{2018}]{sisson+fb18}
Sisson, S.~A., Y.~Fan, and M.~A. Beaumont (Eds.) (2018).
\newblock {\em Handbook of Approximate Bayesian Computation}.
\newblock Chapman {\&} Hall/CRC.

\bibitem[\protect\citeauthoryear{Wood}{Wood}{2010}]{wood10}
Wood, S.~N. (2010).
\newblock Statistical inference for noisy nonlinear ecological dynamic systems.
\newblock {\em Nature\/}~{\em 466\/}(7310), 1102--1104.

\bibitem[\protect\citeauthoryear{Zhu, Ibrahim, and Tang}{Zhu
  et~al.}{2011}]{zhu+it11}
Zhu, H., J.~G. Ibrahim, and N.~Tang (2011).
\newblock {Bayesian influence analysis: a geometric approach}.
\newblock {\em Biometrika\/}~{\em 98\/}(2), 307--323.

\end{thebibliography}

\end{document}